\documentclass[prb,aps,twocolumn,amsmath,amssymb,floatfix,superscriptaddress]{revtex4-1}

\usepackage[dvips]{graphics}
\usepackage{color}

\usepackage{newtxtext,newtxmath}
\usepackage[utf8]{inputenx}
\usepackage{graphicx}
\usepackage{hyperref}
\hypersetup{colorlinks,citecolor=blue,linkcolor=blue}
\usepackage{booktabs}
\usepackage{mathrsfs}
\usepackage[dvipsnames]{xcolor}
\hypersetup{linkcolor=blue,citecolor=blue,urlcolor=blue}
\usepackage[textwidth=17.5cm,textheight=23.5cm,verbose,pdftex]{geometry}
\usepackage{soul}

\usepackage{bm}
\usepackage{amssymb}
\usepackage{textcomp}
\usepackage{gensymb}
\usepackage{comment}
\usepackage{setspace}
\usepackage{tabularx}
\usepackage{cleveref}
\usepackage{tikz}
\usepackage{tikz-3dplot}
\usepackage{placeins}
\usepackage{appendix}
\usepackage[caption=false]{subfig}
\usepackage{cleveref}

\newcommand{\cma}{Co$_2$MnAl}

\begin{document}

\graphicspath{{./figs/}}

\title{Large Violation of the Wiedemann–Franz Law in Heusler, Ferromagnetic, Weyl Semimetal \cma}

\author{Robert A. Robinson}
\affiliation{Department of Physics, Pennsylvania State University, University Park, Pennsylvania, 16802, USA}
\author{Lujin Min}
\affiliation{Department of Materials Science and Engineering, Pennsylvania State University, University Park, Pennsylvania 16802, USA}
\author{Seng Huat Lee}
\affiliation{2D Crystal Consortium, Materials Research Institute, The Pennsylvania State University, University Park, PA 16802, USA}
\author{Peigang Li}
\affiliation{Department of Physics and Engineering Physics, Tulane University, New Orleans, Louisiana 70118}
\author{Yu Wang}
\affiliation{Department of Physics, Pennsylvania State University, University Park, Pennsylvania, 16802, USA}
\affiliation{2D Crystal Consortium, Materials Research Institute, The Pennsylvania State University, University Park, PA 16802, USA}
\author{Jinguo Li}
\affiliation{Superalloys Division, Institute of Metal Research, Chinese Academy of Sciences, 110016 Shenyang, China.}
\author{Zhiqiang Mao}
\email[Email: ]{zim1@psu.edu}
\affiliation{Department of Physics, Pennsylvania State University, University Park, Pennsylvania, 16802, USA}
\affiliation{2D Crystal Consortium, Materials Research Institute, The Pennsylvania State University, University Park, PA 16802, USA}

\date{\today}

\begin{abstract}
The Wiedemann-Franz (WF) law relates the electronic component of the thermal conductivity to the electrical conductivity in metals through the Lorenz number. The WF law has proven to be remarkably robust, however violations have been observed in many topological materials. In this work, we report thermoelectric measurements conducted on Heusler, ferromagnetic, Weyl semimetal \cma\ which shows a drastic, temperature dependent violation of the WF law below 300 K. We then discuss our result in the context of known physical explanations for WF law violation. Both the magnitude and temperature dependence of the violation in \cma\ are extreme, indicating that there may be more than one effect contributing to the violation in this system.
\end{abstract}

\maketitle

\section{Introduction}

Charge carriers in a metal provide for the conduction of both electrical current and heat energy. Since the carrier thermal conductivity, $\kappa_e$, and the electrical conductivity, $\sigma$, are mediated by the same particles, the ratio between them is proportional to the temperature. This proportionality is referred to as the Wiedemann-Franz (WF) law, and is given by $L_0 T=\frac{\kappa_e}{\sigma}$, where $L_0=2.44\times10^{-8}\left[\frac{W \ohm}{K^2}\right]$ is the Lorenz number\cite{He}. The value of the Lorenz number and the WF law have been experimentally verified in a variety of metals and the law has been found to hold over a wide temperature range\cite{Goldsmid}. The law has also been found to be robust to impurity scattering and momentum conserving electron-electron interactions\cite{He,Gooth,Han,Principi}.
 
While the WF law does hold for the majority of metals, violations have been found. In particular, topological materials have been found to be WF violating systems due to their unusual electronic properties\cite{Wakeham,Gooth,Han,Principi,Lee,Jaoui,Zarenia,Xu,Pariari,Rao,Wan,XuCMG}. These violations are of fundamental interest because they indicate that the charge carriers in the material are no longer following classical electron gas behavior; therefore interesting quantum phenomena or strongly correlated behavior may be present in the system. WF violations are also of practical interest because the correlation of thermal and electrical conductivity is a limiting factor in the engineering of high efficiency thermoelectric devices, so deviation from the WF law can open up new engineering possibilities\cite{He}. Violations that increase and decrease the apparent Lorenz number have been observed, with materials such as Li$_{0.9}$Mo$_6$O$_{17}$\cite{Wakeham} and TaP\cite{Han} showing a significant increase, and materials like  Bi$_{2-x}$Cu$_x$Se$_3$\cite{Lee}, WP$_2$\cite{Gooth}, Cd$_3$As$_2$\cite{Pariari}, TaAs$_2$\cite{Rao}, NbAs$_2$\cite{Rao}, and CoSb$_{3-x}$Te$_x$\cite{Wan} showing a decrease.

Understanding the physics behind these violations is a work in progress, but various potential explanations have been proposed. Amongst the proposed explanations, some attribute the breakdown of the WF law directly to the topological nature of the materials while others focus on the fact that the WF law does not hold if the electrons undergo inelastic collisions or are strongly correlated. Examples of explanations that center on the topological nature of the material include non-parabolic band crossings\cite{Wan,Bhandari}, topological Fermi liquid (TFL) behavior\cite{Haldane,Jho}, and axion electrodynamics of the Weyl metallic phase\cite{Kim}. On the other hand, strong screening between electrons leading to many small angle scattering collisions\cite{Jaoui}, large k-space separation of electrons and holes leading to slow recombination of electrons and holes in bulk\cite{Zarenia}, non-Fermi liquid (NFL) behavior\cite{Pariari,Wakeham,Rao}, and the material entering the hydrodynamic regime\cite{Gooth,Principi} are all explanations that depend on complex electron-electron interactions.

While WF law violation has been observed in a number of topological systems, very little work has been done on ferromagnetic Weyl semimetals such as Co$_3$Sn$_2$S$_2$\cite{Morali,Liu,Wang,LiuD} and Co$_2$MnGa\cite{XuCMG,Guin,Belopolski}, which have recently attracted a great deal of interest. Studies on Co$_2$MnGa have found that it contains a Weyl nodal line forming Hopf links\cite{Chang} and that it exhibits large anomalous Hall and anomalous Nernst effects\cite{XuCMG,Guin,Sakai}. The transverse anomalous WF law has also been studied in this system and was found to hold\cite{XuCMG}, however the longitudinal WF law has not been studied in this system.

\begin{figure*}
    \centering
    \subfloat[]{
        \includegraphics[width=0.8\columnwidth]{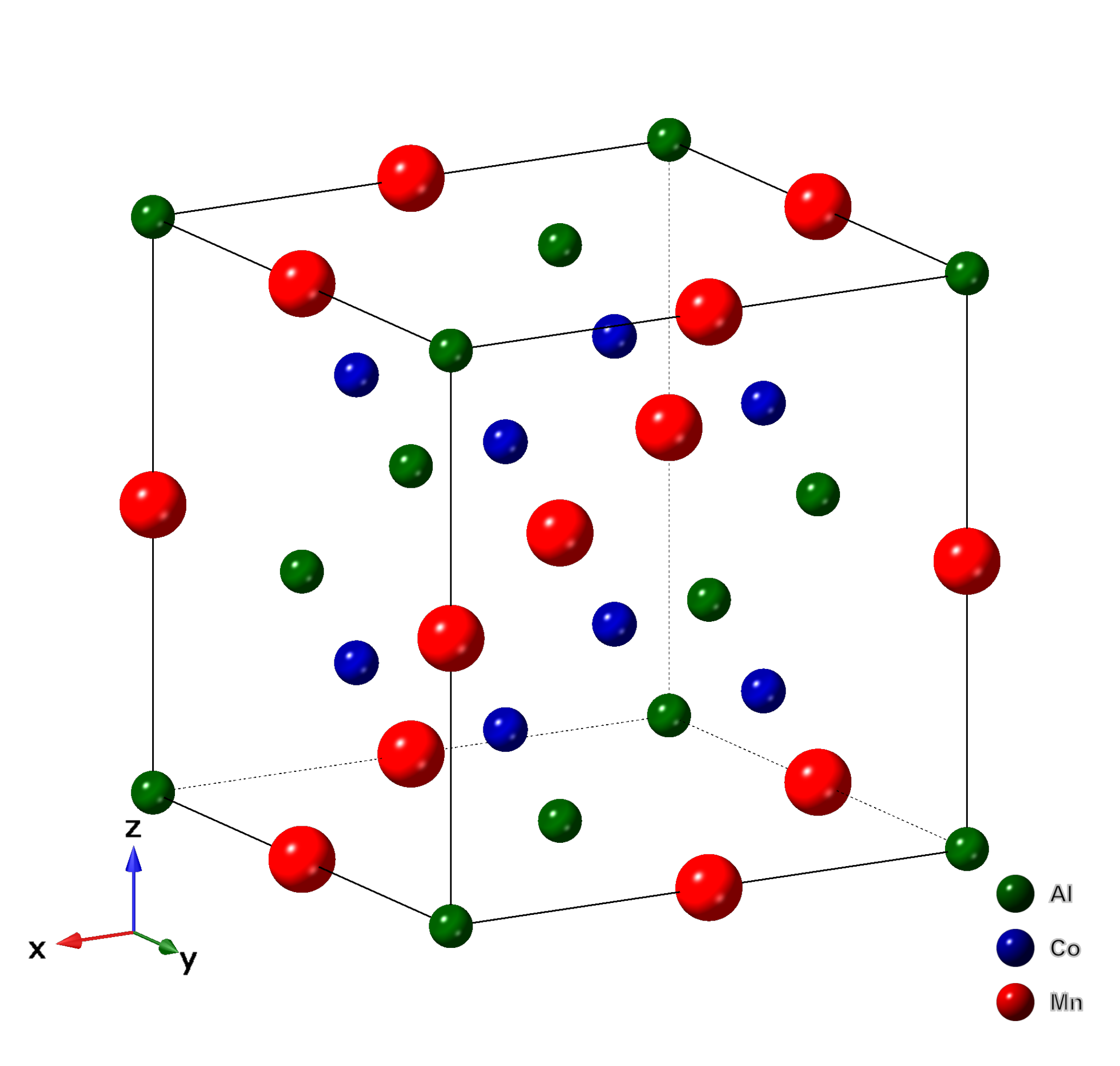}
    }
    \subfloat[]{
        \raisebox{-0.1\height}{\includegraphics[width=\columnwidth]{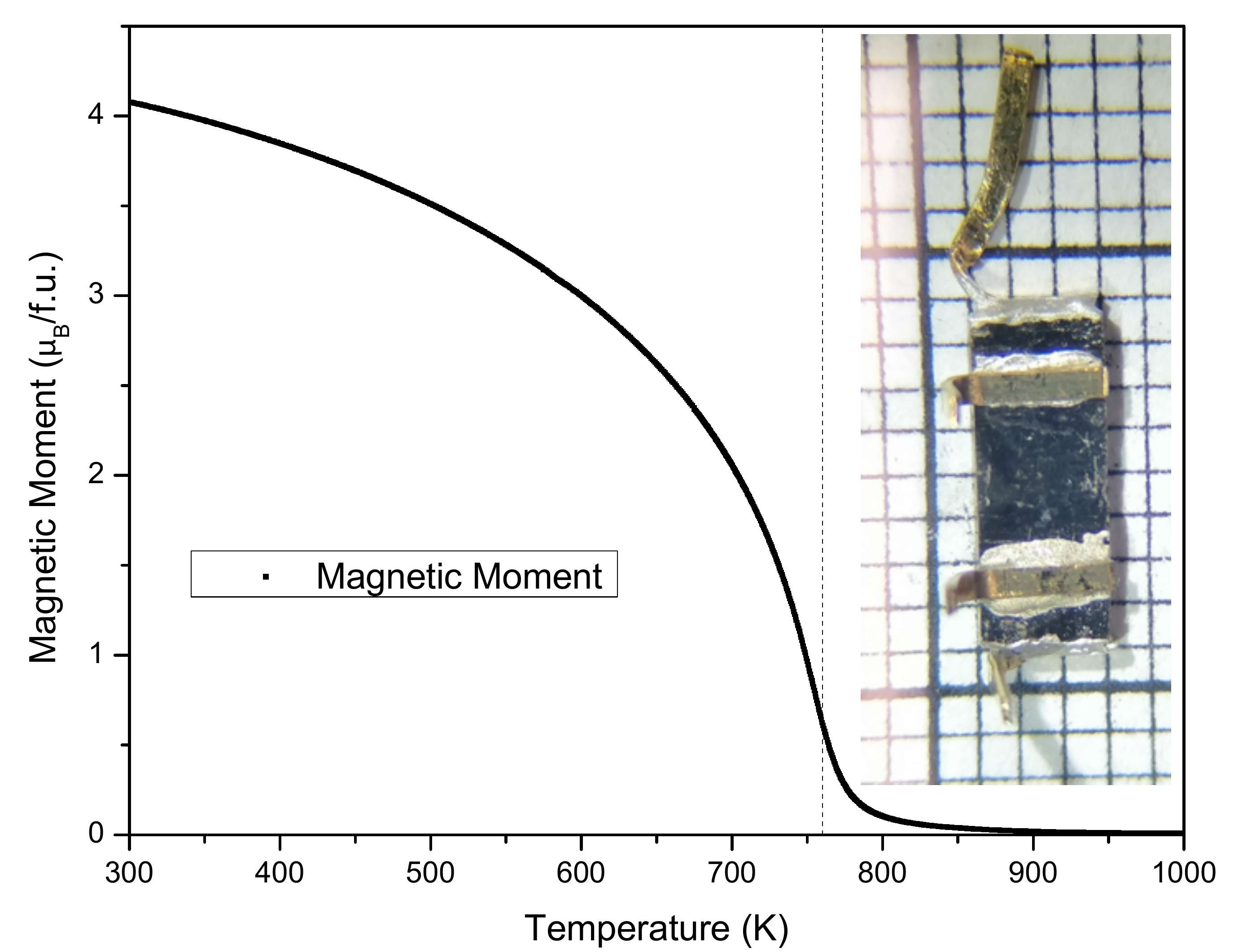}}
    }
    \caption{\textbf{Structure and Sample Characteristics of \cma}} \textbf{(a)} Schematic crystal structure of \cma. \textbf{(b)} Magnetization data for a small piece of single crystal from the same source as the thermal transport sample. The $T_c$ is about 760 K, which is close to the reported value for the $L2_1$ phase\cite{Umetsu}. \textbf{(b, inset)} Picture of the sample used for thermal transport and Laue measurements. Leads are made of gold plated copper and attached using silver epoxy. The small grid squares are 1 x 1 mm$^2$.
    \label{Characterization}
\end{figure*}

Co$_2$MnGa has an isostructural sister compound, \cma, which is also a ferromagnetic Weyl semimetal\cite{LiCMA,Kubler}. Much like Co$_2$MnGa, \cma\ is an ferromagnetic Heusler alloy and has a cubic structure with a space group of $Fm\overline{3}m$ (Fig. \ref{Characterization}a). \cma\ has 2 major structural phases, B2 and L2$_1$\cite{Umetsu}, with the L2$_1$ phase being a ferromagnetic Weyl semimetal\cite{Kubler,LiCMA}. The $L2_1$ phase has a reported Curie temperature of $T_c=726$ K\cite{Umetsu}. What sets \cma\ apart from Co$_2$MnGa is the dependence of its band structure on the applied magnetic field. In \cma, the two lowest conduction bands and 2 highest valence bands cross creating four Weyl nodal rings. These rings form a Hopf-like chain and are protected by the mirror symmetries of the material. When applying a magnetic field along one of the principle crystallographic axes, e.g. the [001] axis, the polarized moments along the [001] direction will cause the mirror symmetry on the $k_x=0$ and $k_y=0$ planes to break gapping out the nodal rings on these two planes creating Weyl nodes. If the magnetic field is applied off axis, all of the nodal rings are gapped out\cite{LiCMA}. This is of particular interest because the gapped nodal rings contribute to the Berry curvature of the material, resulting in a giant room temperature anomalous Hall effect\cite{LiCMA}. This field dependence of the Berry curvature makes \cma\ particularly interesting for experimental study of the effect of Berry curvature tuned by field orientation on transport properties.

We have measured the resistivity, thermal conductivity, and thermopower of \cma\ as a function of temperature and applied field and examined the validity of the WF law for this material. From analysis of the thermal conductivity and resistivity data we observed a drastic violation of the Wiedemann-Franz law as well as nonclassical temperature dependence of the Lorenz number. This breaking of the WF law can be understood in terms of \cma's exotic band structure.

\section{Experimental Method}
\setlength{\parindent}{5ex}

\cma\ single crystals were grown using the floating zone method from rods prepared in an induction furnace. The crystals were annealed for two weeks, one week at 1550 \degree C and another at 873 \degree C\cite{LiCMA}. A crystal plane was then identified and a bar was cut and sanded into a rectangular prism with the dimensions 7.64 x 3.08 x 0.92 mm$^3$ (Fig. \ref{Characterization}b inset). Laue measurements were performed on the sample using a Multiwire Backrelflection Laue and confirmed that the sample is a single crystal. Magnetization measurements were conducted on a small piece from the same source using a Quantum Design SQUID magnetometer from 300-1000 K at 0.5 T.

Resistivity, thermal conductivity, and thermopower measurements were done using a Quantum Design PPMS with a thermal transport option (TTO) probe using a standard 4 lead method. The leads are made of gold plated copper and attached to the sample using silver epoxy (Fig. \ref{Characterization}b inset). Measurements were taken as a function of temperature from 3-300 K at a variety of applied fields from 0-9 T. The applied field was in a direction normal to that of the current and temperature gradient.

At temperatures above 100 K, the machine error in the thermal conductivity data was relatively large (Fig. \ref{ThCond}b). To ensure the reliability of the measured thermal conductivity data above 100 K, statistical significance measurements were carried out\cite{Giunti}. Thermal conductivity was measured repeatedly at a number of fixed temperature and field values, then that data was checked for statistical significance using t-tests with $\alpha=0.05$. The t-tests indicated that the field dependence of the thermal conductivity is statistically significant.

In order to find the Lorenz number of \cma, 2 fits need to be done. The first fit is to thermal conductivity data as a function of applied field at a number of temperatures in order to extract the phonon contribution to the thermal conductivity, $\kappa_{ph}$. The second fit is to the extracted $\kappa_{ph}$ as a function of temperature to generate a function to subtract from the total thermal conductivity. The fit used to construct $\kappa_{ph}(T)$ is only used to subtract from the original thermal conductivity data, so the best mathematical fit was chosen, in this case a Gaussian. The thermal conductivity as a function of temperature has a broad peak near 60 K and the Gaussian is only a good fit for the data at or above the peak, as such the computed Lorenz number $L_0(T)$ is only reported from 100-300 K.

\section{Results and Discussion}
\setlength{\parindent}{5ex}

\subsection{Magnetization and Thermopower}

Magnetization data was collected from 300-1000 K in order to experimentally verify that the sample is in the $L2_1$ phase. \cma\ has 2 major phases, the $L2_1$ phase and the $B2$ phase. According to the literature, the $B2$ phase has a $T_c=677$ K and the $L2_1$ phase has a $T_c=726$ K\cite{Umetsu}. Our sample has a $T_c\approx760$ K, which indicates that it is in the $L2_1$ phase. This is consistent with our early transmission electron microscopy  studies which confirmed the L2$_1$ phase of the floating-zone grown \cma\ crystal\cite{LiCMA}.

\begin{figure}
    \subfloat[]{%
      \includegraphics[width=\columnwidth]{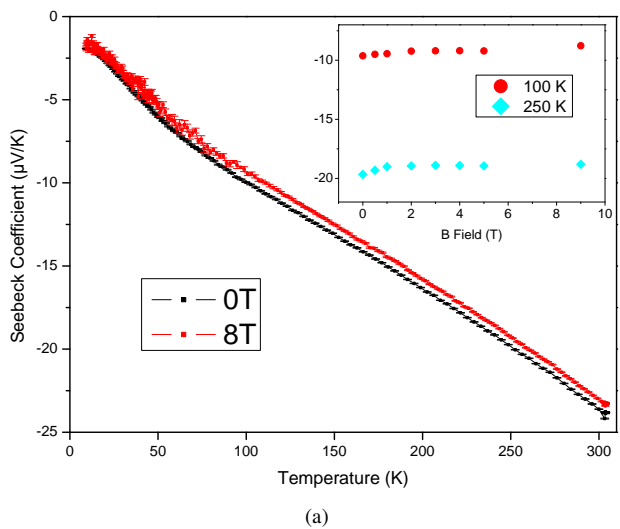}%
    }

    \subfloat[]{%
      \includegraphics[width=\columnwidth]{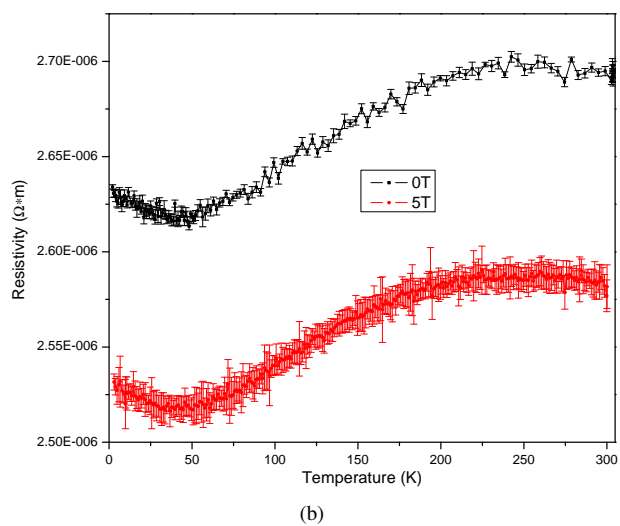}%
    }
\caption{\textbf{(a)} Thermopower as a function of temperature, at 0 and 8 T. The slight overall decrease in magnitude under field is due to the decrease in phonon-magnon scattering. \textbf{Inset:} Thermopower data as a function of field at select temperatures. Above the saturation field of \cma, about 0.5 T, the Seebeck coefficient does not meaningfully depend on the magnitude of the applied field. \textbf{(b)} Resistivity as a function of temperature at 0 and 5T. There is an overall decrease in the magnitude under applied field. Both curves have the same shape which agrees with reported measurements\cite{LiCMA}.}
\label{Seebeck}
\end{figure}

Thermopower measurements were collected from 2-300 K at fixed fields from 0-9 T (Fig. \ref{Seebeck}a), and the value of the Seebeck coefficient reaches a maximum of almost -25 $\left[\frac{\mu V}{K}\right]$ at 300 K and has a similar magnitude to that of \cma's sister compound Co$_2$MnGa\cite{Guin}. Since measurements below 25 K under high magnetic field require high precision calibration which has not been completed in our system, we will constrain our analysis to data collected at 50 K or above. The Seebeck coefficient is almost linear throughout the temperature range, which is in good agreement with expected metallic behavior\cite{Goldsmid,Behnia}. Over the whole temperature range the value of the Seebeck coefficient is negative indicating that electrons are the majority charge carriers in \cma. The Seebeck coefficient is not suppressed at high temperatures indicating an absence of the bipolar effect\cite{Guin}. Both the n-type behavior and lack of a bipolar effect are in agreement with the calculated band structure which shows large electron pockets and small hole pockets\cite{Kubler,LiCMA}.

There is a slight decrease in the magnitude of the Seebeck coefficient as the magnetic field is applied (Fig. \ref{Seebeck}a inset). This decrease is caused by the fact the phonon-magnon scattering is suppressed under applied field, and that reduction in scattering leads to a larger contribution to thermal conductivity from phonons which in turn decreases the thermopower. Given that \cma\ has a saturation field below 0.5 T\cite{LiCMA} it is not surprising that above 0.5 T further increasing the field has no substantial impact on the thermopower measurement (Fig. \ref{Seebeck}a inset).

\subsection{Thermal Conductivity}

\begin{figure}
    \subfloat[]{%
      \includegraphics[width=\columnwidth]{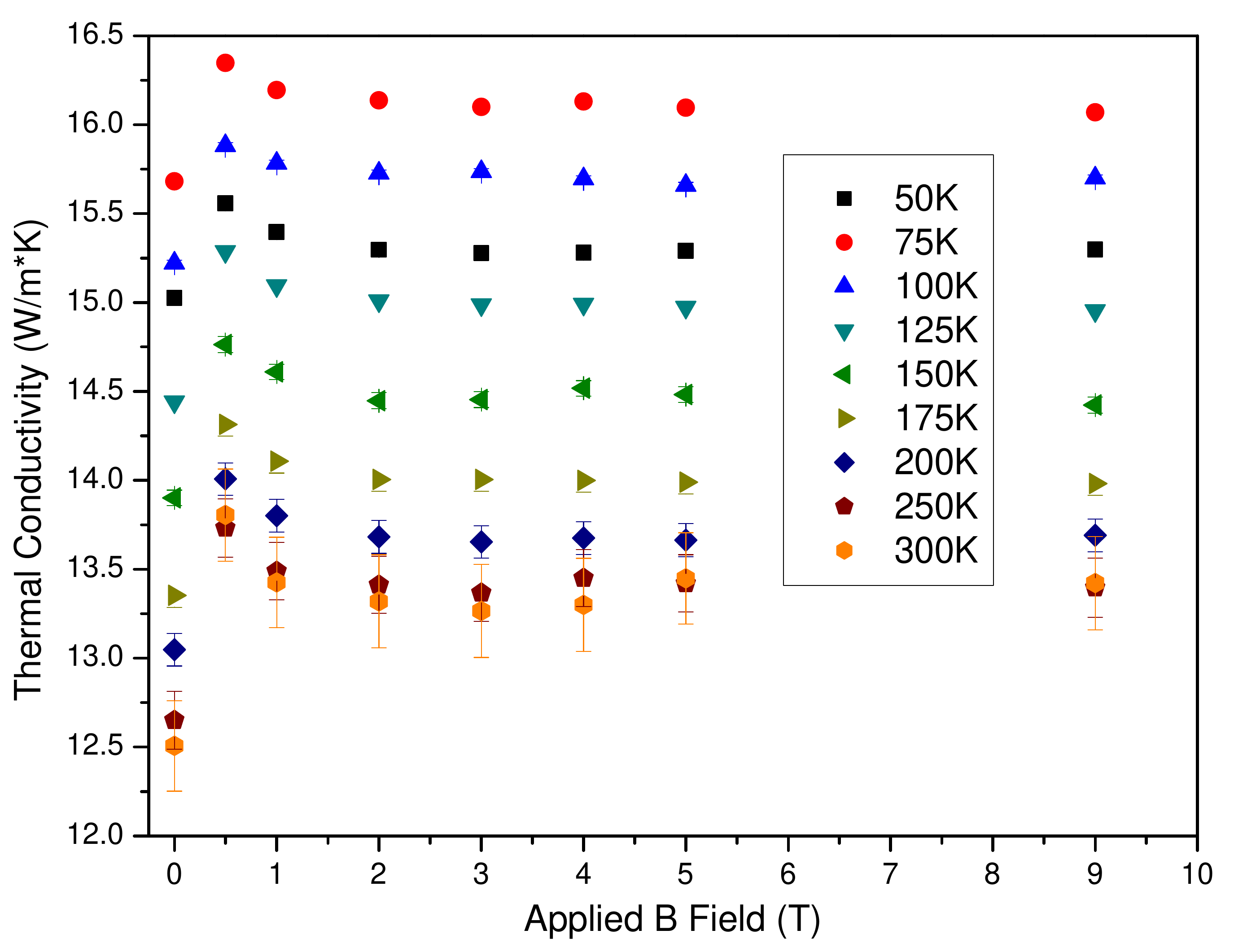}%
    }

    \subfloat[]{%
      \includegraphics[width=\columnwidth]{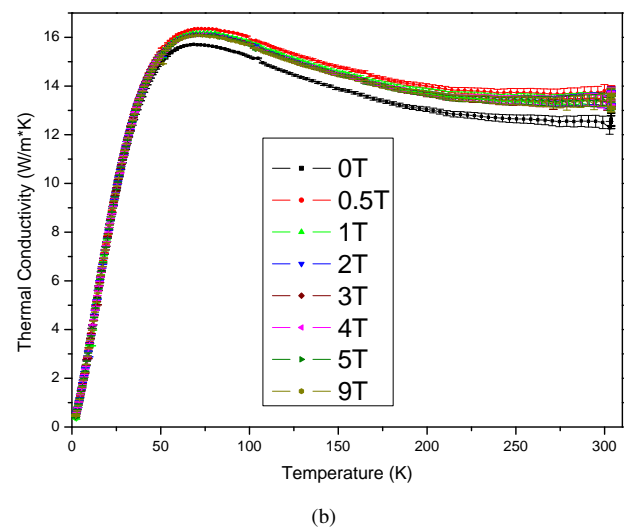}%
    }
\caption{\textbf{(a)} Thermal conductivity vs. applied magnetic field at various temperatures. The initial jump in the magnitude of the thermal conductivity is due to the suppression of phonon-magnon scattering by the applied field. The gradual decay above 0.5 T is due to the suppression of the electronic component of the thermal conductivity by the applied field. All data includes error bars, but below 150 K the error is sufficiently small that it is hidden in the data point. \textbf{(b)} Thermal conductivity vs. temperature. The behavior of thermal conductivity as a function of temperature is in good agreement with similar materials and shows a slight increase with decreasing temperature up to a peak at around 60 K after which it drops sharply to zero as the phonon degrees of freedom are frozen out.}
\label{ThCond}
\end{figure}

Thermal conductivity data was collected from 2-300 K at fixed field values from 0-9 T. The magnetic field was applied in a direction normal to the applied current and temperature gradients. When going from 0 to 0.5 T there is a pronounced increase in the magnitude of the thermal conductivity at all temperatures (Fig. \ref{ThCond}a). This increase has to do with the ferromagnetic nature of \cma\ and phonon-magnon coupling. In magnetic materials, phonons can interact with spin-wave fluctuations, magnons, and this interaction scatters the phonons decreasing their mean free path. As an external field is applied, the magnetic moments in the material becomes more ordered and magnons are suppressed. Magnon suppression leads to less phonon scattering allowing the phonons to more efficiently move thermal energy through the material; this leads to an increase in the thermal conductivity up to the saturation field of the system\cite{Tsujii}.

Once the applied field is above the saturation field of \cma, there is a clear decay of the thermal conductivity (Fig. \ref{ThCond}a). This decay is due to the suppression of $\kappa_e$ by the applied field. This decay is caused by induced cyclotron motion limiting the ability of the electrons to travel through the material. As the field is increased, the magnitude of the Lorentz force on the electrons increases until their contribution to the thermal conductivity is almost entirely suppressed\cite{Han}. From the magnitude of this suppression, it is clear that the thermal conductivity of \cma\ is phonon dominated, especially when compared with metals like Cu in which the thermal conductivity is dominated by charge carriers\cite{Hunsicker}. In order to study the WF law, the phonon and electron contributions need to be separated as the WF law depends only on the electron contribution. We separated the two components using a procedure similar to that conducted by Han et al.\cite{Han}.

\begin{figure}
    \subfloat[]{%
      \includegraphics[width=\columnwidth]{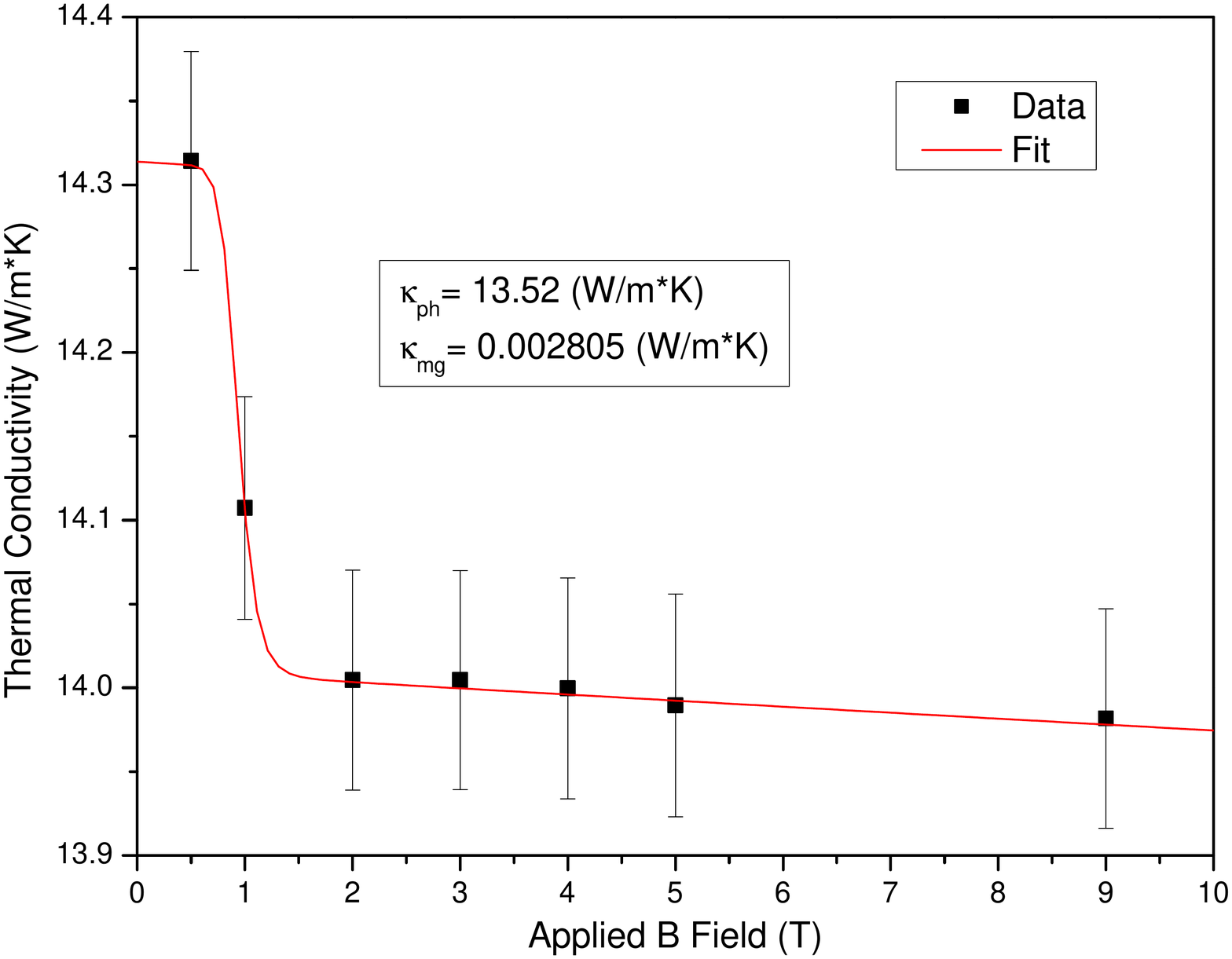}%
    }

    \subfloat[]{
      \includegraphics[width=\columnwidth]{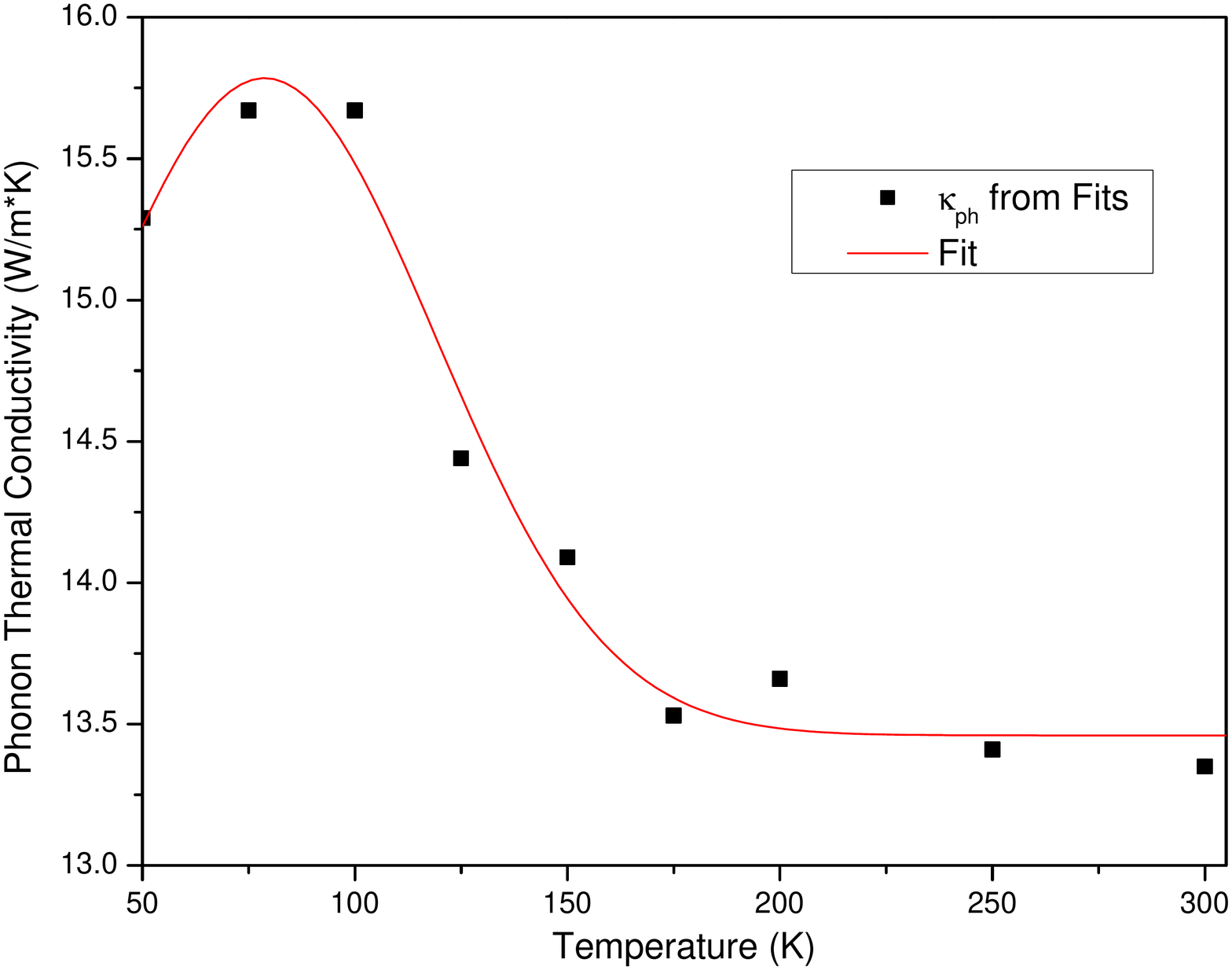}%
    }
\caption{\textbf{(a)} An example of the fit of the field dependence of the thermal conductivity to eq. \ref{ThCond_eqn} at 175 K. Fitting results for $\kappa_{ph}$ and $\kappa_{mg}$ are shown. \textbf{(b)} Fit of calculated $\kappa_{ph}$ values to a Gaussian equation in order to subtract that term from the total thermal conductivity.}
\label{Fit_ThCond}
\end{figure}

The process for fitting the data involves several steps, the first of which is fitting the thermal conductivity vs. applied field data at a variety of fixed temperatures to\cite{Han,Lau}

\begin{equation}
    \kappa(T,B)=\kappa_{ph}(T)+\frac{\kappa_e(T,B=0)}{1+\beta_e(T)B^m}+\kappa_{mg}Te^{\frac{-g\mu_B B}{k_B T}}
    \label{ThCond_eqn}
\end{equation}

\noindent where $\kappa_{ph}\ \&\ \kappa_e$ are the contributions from the phonons and electrons to the thermal conductivity respectively, $\kappa_{mg}Te^{\frac{-g\mu_B B}{k_B T}}$ is the contribution to the thermal conductivity from magnons\cite{Lau}, $\beta_e(T)$ is proportional to the zero-field mean free path of the electrons, $m$ is related to how the electrons scatter\cite{Han}, $g$ is the g factor\cite{Lau}, and $\mu_B$ is the Bohr magneton. The goal of this fit is to extract the values of $k_{ph}$ at a number of temperatures so that an overall temperature dependence can be determined. Fig. \ref{Fit_ThCond}a shows an example data and fit. From these fits, we find that the contribution due to magnons is negligible, which is not surprising given that the magnetization of the material saturates at very low field (0.5 T for the field along the spin easy axis), meaning that the phonon scattering by magnons is strongly suppressed in the entire field range of the fit. We will drop the magnon contribution for the rest of our analysis because it is negligible. It is also worth noting that we do not include the 0 T data in these fits as the phonon-magnon coupling at 0 T is not accounted for in eq. \ref{ThCond_eqn}, so cannot be fit by this method.

Once we have $\kappa_{ph}$ at a number of temperatures, that data is plotted and fit to a mathematical equation in order to determine a temperature dependence over the entire temperature range (Fig. \ref{Fit_ThCond}b). The equation for $\kappa_{ph}$ can then be subtracted from the total measured thermal conductivity in order to determine $\kappa_e$. $\kappa_e$ and $\sigma$ can then be plugged into

\begin{equation}
    L(T)=\frac{\kappa_e}{\sigma T}
    \label{WF Law}
\end{equation}

\noindent in order to find the temperature dependent Lorenz number of the system.

\subsection{Lorenz Number}

\begin{figure}
    \subfloat[]{%
      \includegraphics[width=\columnwidth]{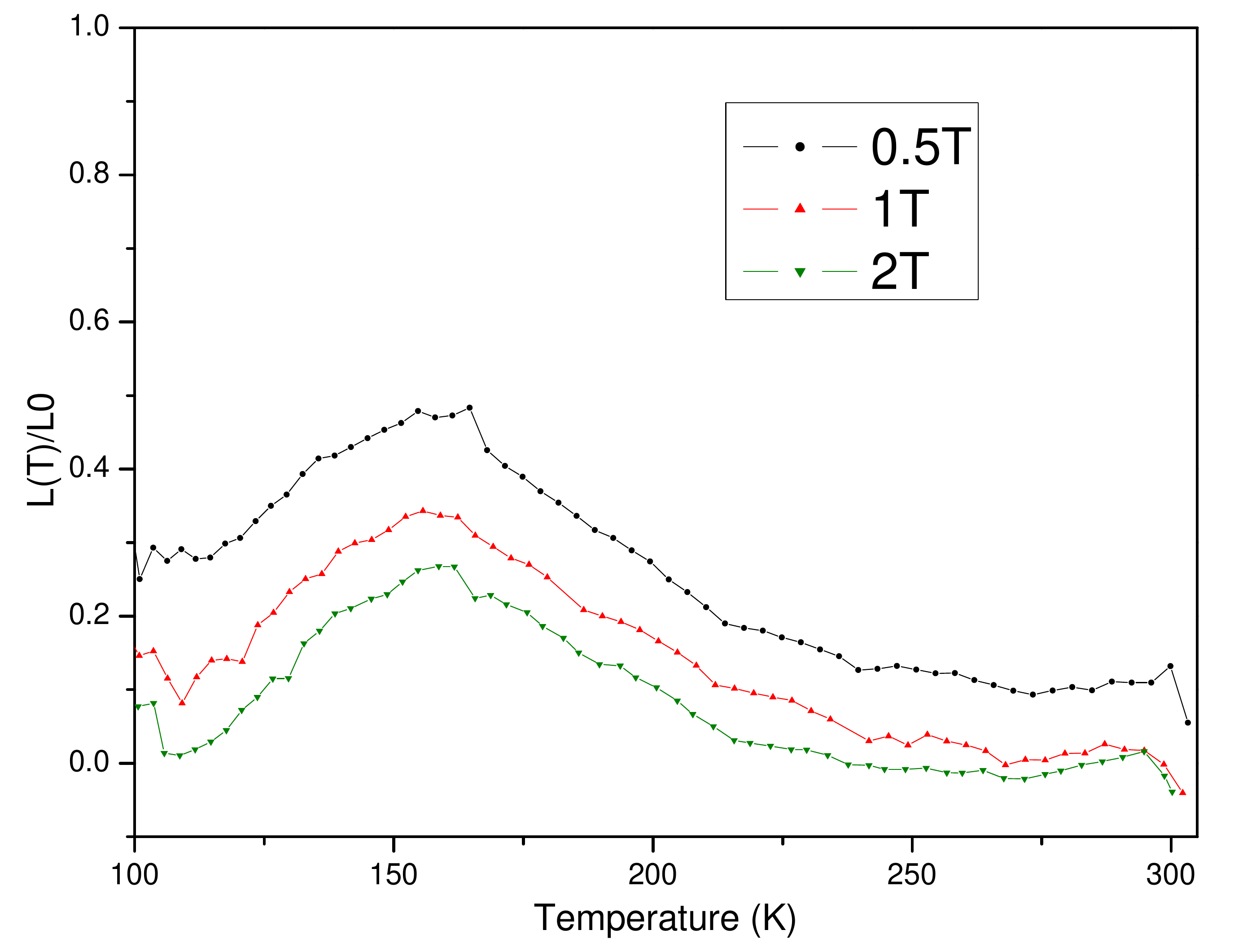}%
    }

    \subfloat[]{%
      \includegraphics[width=\columnwidth]{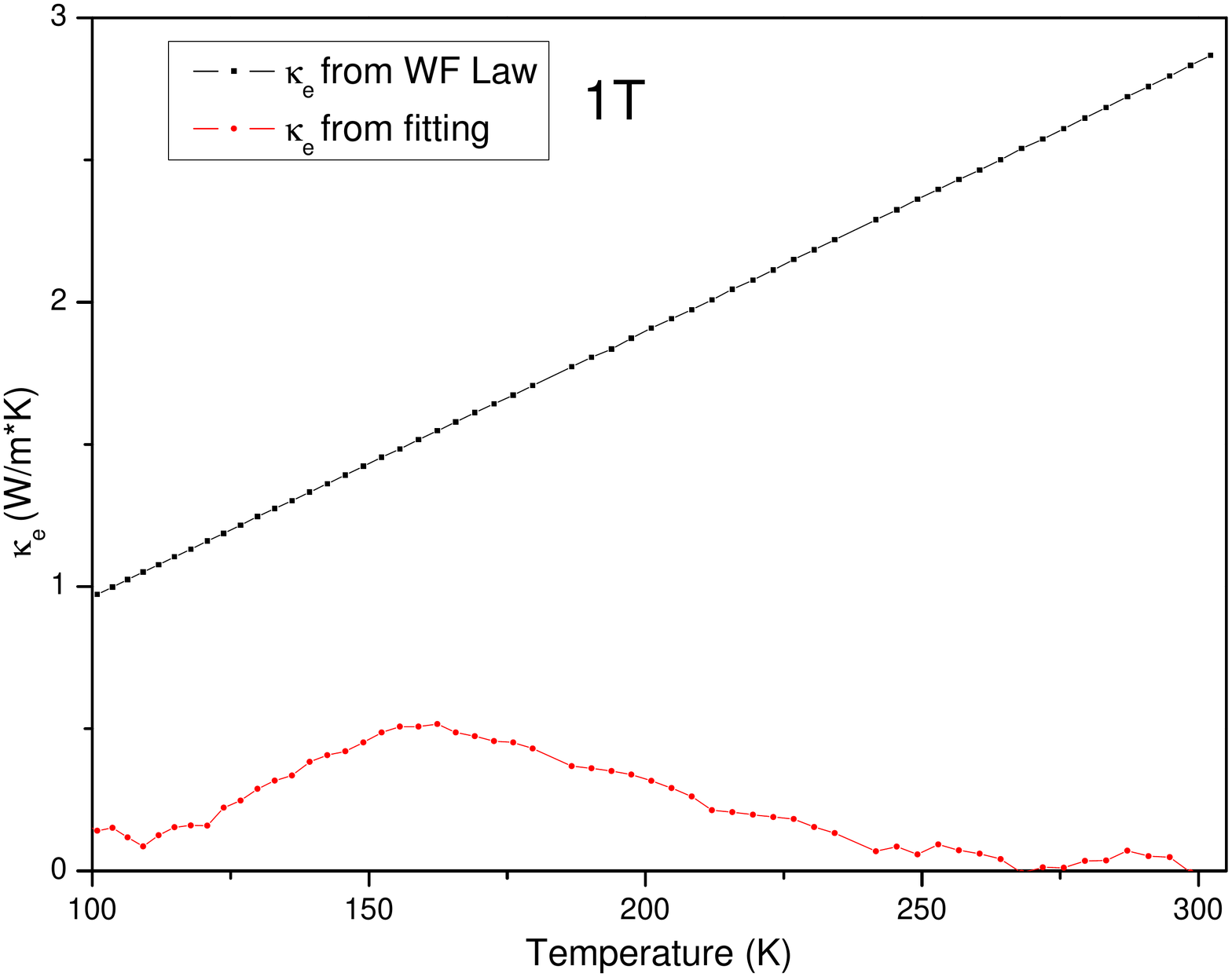}%
    }
\caption{\textbf{(a)} Normalized Lorenz number as a function of temperature at 0.5, 1, \& 2 T. The calculated value becomes arbitrarily small around 300 K and 100 K with a peak of about 0.4 near 160 K. This unusual temperature dependence could indicate that more than one effect is contributing to the violation. \textbf{(b)} comparison of $\kappa_e$ predicted by the Wiedemann–Franz law (black) and the results of our measured thermal conductivity minus calculated $\kappa_{ph}$. The calculated $\kappa_e$ is clearly much lower than the one predicted by the Wiedemann–Franz law, further indicating that there is a violation.}
\label{L0}
\end{figure}

Our results for $\frac{L(T)}{L_0}$ are presented in Fig. \ref{L0}a, there is clearly both a drastic departure from the classical Lorenz number, and a clear nonlinear temperature dependence. The value of the Lorenz number is reduced throughout the entire temperature range, with the closest value being about 40\% of $L_0$ near 155 K, then becoming arbitrarily small at both low temperatures and temperatures near 300 K. The overall decrease in the Lorenz number indicates that either the electrons are contributing substantially less to the conduction of heat vs. charge or that some other mechanism is enhancing the charge conduction. The temperature dependence implies that whatever physical phenomena are causing this deviation are dependent on the temperature. 
 
Many possible theories have been proposed for WF law violating behavior as noted above; some of them can apply to \cma\ while others cannot. Tomonaga-Luttinger liquid behavior\cite{Wakeham}, topological Fermi liquid behavior\cite{Haldane,Jho}, axion electrodynamics of the Weyl metallic phase\cite{Kim}, non-Fermi liquid behavior\cite{Pariari,Wakeham,Rao}, and chiral zero sound behavior\cite{Xiang} could, in general, lead to WF law violation. However, none of these theories can explain our results for \cma\ because either their observation requires experiments we did not conduct, they require the material has properties not present in \cma, or the behavior they cause is not consistent with the results of our measurements. With that point addressed, we will now discuss those explanations that could apply to \cma\ and how they relate to the observed WF law violation.
 
One possible cause of WF violation is that the WF law and Lorenz number are derived under the assumption that the material has a simple parabolic band structure\cite{Wan,Bhandari}. However, this is clearly not the case in \cma, because it is a Weyl semimetal; the band crossings near the Fermi energy are linear and form Weyl nodal lines\cite{Kubler}. Through theoretical calculations, Wan et al.\cite{Wan} demonstrated that such non-parabolic band structure can lead to substantial discrepancies in the value of the Lorenz number. In that work, non-parabolic band structure was used to explain the WF violation in CoSb$_{3-x}$Te$_x$ near and above room temperature. The effect that a non-parabolic band structure has on the Lorenz number should be uniform over the entire temperature range, so it could contribute to the overall decrease in magnitude of $L_0$ observed in \cma, but cannot explain the temperature dependence.

A second possible explanation is slow recombination of electrons and holes. This slow recombination has been predicted in type-II Weyl semimetals and is due to a large k space separation between the electrons and holes\cite{Zarenia}. Proposed by Zarenia et al.\cite{Zarenia}, this theory states that classical thermal conductivity also contains a term from the recombination of electrons and holes in the bulk of the material. This would mean that the classical Lorenz number also depends on this term. However, if electrons and holes are sufficiently separated in k space, as they are in some Weyl semimetals, then this recombination process slows so much that its contribution to the thermal conductivity drops out, leading to a lower overall Lorenz number. In their work, they propose this mechanism as a possible explanation for the WF law violation observed by others\cite{Jaoui,Gooth} in WP$_2$\cite{Zarenia}. \cma\ has both electron and hole pockets with the largest hole pocket being at the center of the Brillouin zone and the largest electron pocket being at the zone boundary\cite{Kubler,LiCMA}. The electron pockets are substantially larger than the hole pockets, but the fact that both are present and separated means that \cma\ could exhibit slow recombination of electrons and holes. Therefore, this effect could contribute to the overall decrease in magnitude of the Lorenz number over the entire temperature range in \cma.

Strong screening between electrons leading to many small angle scattering collisions\cite{Jaoui} could also play a part in the WF violation observed in \cma. This small angle scattering is proposed to take place between electrons and holes and could either be Umklapp or inter-band in nature. This behavior can also lead to the material entering the hydrodynamic regime\cite{Jaoui}. The hydrodynamic regime is one in which the electrons are strongly interacting and energy dissipation most strongly depends on many small angle electron-phonon or momentum conserving electron-electron scattering events\cite{Pariari,Gooth}. In the hydrodynamic regime, the electron-electron scattering time is substantially shorter than the electron-phonon and electron-impurity scattering times. Charge currents can only be relaxed by momentum nonconserving processes, while thermal currents can be relaxed by momentum conserving interactions, like electron-electron scattering. So in the hydrodynamic regime where electron-electron scattering happens much more frequently, the thermal current relaxation time is significantly shorter than the charge current relaxation time\cite{Pariari}. This mismatch can allow for violation of the WF law, which according to theory\cite{Pariari} and experiment\cite{Gooth} can lead to an arbitrarily small Lorenz number in the hydrodynamic limit. In \cma, the band structure near the Fermi energy has a large contribution due to the 3d orbitals of Co\cite{Kubler,LiCMA} and it is generally expected that when 3d orbitals are involved, correlated electron behavior can be present meaning that strong screening can occur. Further, Co$_2$MnAl contains Weyl nodal lines forming Hopf-links\cite{Guin} as noted above. There is evidence that a nodal line structure can lead to strongly correlated electron behavior\cite{Shao}. It follows that, the nodal lines in \cma\ : may also lead to correlated behavior and strong screening between electrons. The theory predicts that in order to observe WF law violation due to strong screening between electrons the material needs a component of the Fermi surface located at the zone boundary, which is the case in \cma\cite{Kubler,LiCMA}. According to DFT calcutaions done by Jaoui et al.\cite{Jaoui}, this behavior can lead to a substantial decrease in the Lorenz number of a material, and has been observed in W\cite{Wagner} and proposed as an explanation for the WF violation observed in WP$_2$\cite{Gooth}. If the hydrodynamic behavior is present in \cma, it would serve as an explanation for the arbitrarily small Lorenz number observed at low temperature. The theory also predicts that the thermal conductivity should have a broad maximum when the momentum relaxation time and the electron-electron interaction contribution to the thermal relaxation time are equal\cite{Principi}. This may help to explain the temperature dependence and maximum observed in the Lorenz number of \cma\ (Fig. \ref{L0}a).

As we have discussed, a variety of mechanisms could contribute to the WF law violation observed in \cma. However, no one mechanism discussed here can satisfactorily explain the entirety of the behavior; further experimental work is needed in order to elucidate the origins of the WF law violation. We suggest that measurements be done to determine whether \cma\ exhibits strongly correlated electron behavior. Optical and magneto-optical spectroscopy measurements, like those performed on ZrSiSe\cite{Shao}, could be used to probe \cma\ for nodal lines as well as correlated behavior due to those nodal lines. ARPES\cite{Damascelli} and specific heat\cite{Naren} measurements are also capable of detecting strongly correlated electron behavior, and so could help to better understand the underlying physics present in \cma.

The fitting we did to find $\kappa_e$ is not the most common method, usually the WF law is used to compute $\kappa_e$ from $\sigma$\cite{Das}. As a method of confirming WF law violation in \cma, we computed $\kappa_e$ from eq. \ref{WF Law} using our electrical resistivity data (Fig. \ref{Seebeck}b) assuming the classical Lorenz number and compared that value to the one calculated by our fits (Fig. \ref{L0}b). Using the WF law, we would expect $\kappa_e$ to be over 1 $\left[\frac{W}{m*K}\right]$ and have linear temperature dependence, which clearly does not agree with our computed $\kappa_e$ (Fig. \ref{L0}b). Further, if the WF law held in \cma, then we would expect a substantially larger suppression of the thermal conductivity due to applied field than what we observed (Fig. \ref{ThCond}a), this discrepancy serves as further confirmation of a WF law violation.

 We would also like to briefly address the anomalous transverse WF law, a WF law behavior induced by the anomalous Nernst effect in a direction normal to the current and temperature gradient. Like the WF law, the anomalous transverse WF law has been studied in some Weyl semimetals, and some, like Mn$_3$Ge\cite{Xu} have shown a violation, while others, like Co$_2$MnGa\cite{XuCMG,Guin} have not. The anomalous transverse WF law violation in Mn$_3$Ge is attributed to a large energy dependence of the Berry curvature near the Fermi energy\cite{Xu}. The fact that there was no anomalous transverse WF violation in Co$_2$MnGa does not conflict with the findings of our work as the mechanism of the WF law in that case is the anomalous Nernst effect\cite{XuCMG}, which is not relevant to the measurements we conducted.

\section{Summary}
\setlength{\parindent}{5ex}

Resistivity, thermopower, and thermal conductivity were measured on a \cma\ single crystal in the $L2_1$ phase. Through our measurements we were able to determine that the thermal conductivity is phonon dominated. By analyzing the collected data, a drastic, temperature dependent WF law violation was observed. We addressed potential explanations of WF violating behavior and found that while no one explanation is sufficient to explain the entirety of the behavior, non-parabolic band crossings, large k-space separation of electrons and holes leading to slow recombination of electrons and holes in bulk, strong screening between electrons leading to many small angle scattering collisions, and the material entering the hydrodynamic regime could all possibly play a part in the observed violation. We conclude that further experimental work such as optical spectroscopy, ARPES and heat capacity measurements is needed in order to determine the exact mechanisms involved in \cma. 

\section{Acknowledgments}
\setlength{\parindent}{6ex}

 This work is supported by the US National Science Foundation under grants DMR 1917579 and 1832031.

\section{Data availability}
\setlength{\parindent}{6ex}

Data is available upon request from author.

\bibliographystyle{iopart-num}
\bibliography{biblio}
\end{document}